\documentclass[aps,prl,twocolumn,showpacs,superscriptaddress]{revtex4}

\usepackage{graphicx}
\usepackage{amsmath}
\usepackage{amssymb}

\input{epsf}
\begin{document}

\title{ Ultracold Rb-OH collisions and prospects for sympathetic
cooling }

\author{Manuel Lara}
\affiliation{JILA, NIST and Department of Physics, University of Colorado,
Boulder, CO 80309-0440}
\author{John L. Bohn}
\affiliation{JILA, NIST and Department of Physics, University of Colorado,
Boulder, CO 80309-0440}
\author{Daniel Potter}
\affiliation{Department of Chemistry, University of Durham,
South Road, Durham, DH1 3LE, United Kingdom}
\author{Pavel Sold\'{a}n}
\affiliation{Doppler Institute, Department of Physics, Faculty of
Nuclear Sciences and Physical Engineering, Czech Technical University,
B\v{r}ehov\'{a} 7, 115 19 Praha 1, Czech Republic}
\author{Jeremy Hutson}
\affiliation{Department of Chemistry, University of Durham,
South Road, Durham, DH1 3LE, United Kingdom}

\date{\today}

\begin{abstract}
We have computed {\it ab initio} cross sections for cold collisions
of Rb atoms with OH radicals.  We predict collision rate constants
of order $10^{-11}$ cm$^3$/s at temperatures in the range 10-100 mK
at which molecules have already been produced. However, we also find
that in these collisions the molecules have a strong propensity for
changing their internal state, which could make sympathetic cooling
of OH in a Rb buffer gas problematic in magnetostatic or
electrostatic traps.
\end{abstract}


\maketitle

A great way to make something cold is to place it in thermal contact
with something else even colder.  Thus, for many years, it has been
possible to cool one species of ion sympathetically, by placing it
in contact with another species that is being actively cooled
\cite{Drullinger,Larson}.  More recently, ultracold neutral atoms
have also been sympathetically cooled \cite{Myatt}. At slightly
higher temperatures, atoms and molecules have been sympathetically
cooled in a helium buffer gas \cite{deCarvalho}. Molecules are
widely regarded as worth cooling to ultralow temperatures (typically
around 1$\mu$K), where de Broglie wavelengths become large, and
intermolecular interactions are dominated by long-range forces.
Under such circumstances, the possibility of permanent electric
dipole moments is expected to lead to novelties in such areas as
precision measurement, collisions and chemistry, quantum degenerate
matter, and quantum information theory \cite{EPJD}.

A host of molecular cooling techniques have been proposed
\cite{EPJD}, but two that stand out are buffer-gas cooling (BGC), in
which the cold He gas cools the molecules, and Stark deceleration
(SD), in which polar molecules are slowed by carefully designed
time-varying electric fields \cite{vanderMeerakker}. These
techniques tend to produce ``lukewarm'' gases, with translational
temperatures in the 10-100 mK range. They are therefore somewhat at
a disadvantage with respect to the direct production of molecules by
photoassociation from a gas of ultracold atoms \cite{Jones06}.
Nevertheless, BGC and SD are extremely appealing for the far greater
variety of molecular species that they can cool.

In this Letter we are concerned with the possibility of turning a 10
mK gas of molecules into a 10 $\mu$K gas, by sympathetically cooling
with Rb.  For this cooling to be effective, the cross sections must
be favorable.  Namely, elastic scattering should occur frequently
(with large cross sections $\sigma_{\rm el}$), to make the necessary
thermal contact.  On the other hand, inelastic collisions that
change the state of the molecules should occur infrequently (with
small cross sections $\sigma_{\rm inel}$). This is because either
magnetostatic or electrostatic trapping demands that the molecules
remain in a well-defined internal state.  Until now, cross sections
for $m$-level-changing cold collisions of alkali atoms with molecules were
wholly unknown. In this Letter we present the first such collision
results, using a set of complete {\it ab initio} potential energy
surfaces (PES's), and incorporating the hyperfine structure of the
collision partners.

Our prototype system for this study consists of OH molecules
in a bath of Rb atoms.  This is  a particularly
timely example:  the OH radical has been successfully
slowed by Stark deceleration techniques in at least two laboratories,
which can now produce $\sim 10$ mK packets of these molecules
on demand \cite{vanderMeerakker,Bochinski}.
At the same time, Rb is easily cooled and trapped in
copious quantities at tens or hundreds of $\mu$K,
making it an ideal target for the molecules.  In addition,
the Rb-OH collision system
is subject to a ``harpooning'' process, similar to
that in Rb-NH \cite{Sol04}, where during a collision the valence electron
jumps from Rb to OH \cite{Levine}.  This process is without precedent in the
cold collisions literature.  Thus, even apart from sympathetic
cooling, the Rb-OH scattering system is quite rich from a cold
collisions perspective.

We begin by describing the potential energy surfaces (PES's)
of the system, as  functions of $(R,\theta)$, where $R$ is the
distance between the atom and diatom, and $\theta$ is the angle
that the Jacobi vector ${\vec R}$ makes with  the OH axis.
The OH monomer has a $^2\Pi$ ground state arising from a $\pi^3$
configuration, while the ground state of Rb is $^2S$. This produces
$^1\Pi$ and $^3\Pi$ states for linear RbOH, which split into $^1A'$,
$^1A^{\prime\prime}$ and $^3A'$, $^3A^{\prime\prime}$ surfaces at
nonlinear geometries. We will refer to these as the covalent states.
In addition, RbOH has an ion-pair state analogous to the ones
previously found for RbNH \cite{Sol04}. The ion-pair threshold for
Rb$^+$ ($^1S$) + OH$^-$ ($^1\Sigma^+$) lies only 2.35 eV above the
ground state at $R=\infty$ and produces a $^1\Sigma^+$ ($^1A'$)
state that cuts steeply down, crossing the covalent states near 6
\AA. This is an actual crossing at linear geometries, where the
ion-pair and covalent states have different symmetries, but there
are avoided crossings between the two states of $^1A'$ symmetry at
nonlinear geometries. There is therefore a conical intersection
between the two $^1A'$ surfaces at linear geometries.

Potential energy surfaces for all 5 electronic states of RbOH were
calculated by complete active space self-consistent field (CASSCF)
calculations followed by multireference configuration interaction
(MRCI). A full description will be given elsewhere \cite{fullpaper}.
All calculations used the MOLPRO package \cite{MOLPRO}, with
aug-cc-pVTZ basis sets \cite{Dunning} in uncontracted form for O and
H and the ECP28MWB small-core quasirelativistic effective core
potential \cite{ECPs} for Rb, with the valence basis set from Ref.\
\cite{Sol03}. The CASSCF calculation included all configurations
that can be formed from the lowest (10,3) orbitals of
($a',a^{\prime\prime}$) symmetry, with the lowest (5,1) orbitals
doubly occupied. Energies were calculated in Jacobi coordinates
($R,\theta)$ on a grid of 25 unequally spaced points in $R$ from 2
to 12 \AA\ and 11 Gauss-Lobatto quadrature points in $\theta$. The
two $^1A'$ surfaces were transformed to obtain two quasidiabatic
diagonal surfaces and a coupling surface, using a mixing angle
derived from matrix elements of the OH $\hat L_z$ operator. Finally,
the quasidiabatic covalent energies were extrapolated to $R=\infty$
using $C_6$ and $C_7$ coefficients obtained from coupled cluster
(CCSD) calculations at $R=15$, 25 and 100 \AA. The singlet and
triplet sum and difference surfaces were expanded in normalized
associated Legendre functions and the radial coefficients were
interpolated using the reciprocal-power reproducing kernel Hilbert
space (RP-RKHS) method \cite{Ho96,Sol00}.

The resulting diabatic covalent surfaces have wells $337$ and $511$
cm$^{-1}$ deep at linear Rb-OH geometries for the singlet and
triplet surfaces respectively. For the $A'$ surfaces this linear
well is the absolute minimum, but the $A^{\prime\prime}$ surfaces
have absolute minima at non-linear geometries ($\theta\approx
125^\circ)$ with depths of $405$ and $615$ cm$^{-1}$ respectively.
The ion-pair state is very much deeper, with a well $\sim 26000$
cm$^{-1}$ below the neutral threshold.

To perform scattering calculations on these surfaces, we expand
the wave function into an appropriate basis set.  This set will
consist of the quantum numbers of the atom and diatom in the
separated limit.  For the atom, electronic ($s_a$) and nuclear
($i_a$) spins are coupled to make total spin $|f_a m_{f_a} \rangle$
in the lab frame.  For the diatom, the electronic wave function
is specified in Hund's coupling case (a) by
the total electronic angular momentum $j$, with projections
$m$ in the lab frame and $\omega$ in the frame rotating with the
molecule. As usual, $\omega$ is separated into its orbital
and spin components, $\omega = \lambda + \sigma$.
 Linear combinations
$|{\bar \lambda} {\bar \omega} \epsilon \rangle =
(|\lambda \omega \rangle + \epsilon |-\lambda -\omega\rangle) /\sqrt{2}$
are constructed to produce states of good parity  appropriate
to the zero-electric-field case we consider here.  Finally,
$j$ is coupled  to the nuclear spin $i_d$ of the hydrogen atom
to yield total diatom spin $|f_dm_{f_d} \rangle$.  The partial
wave quantum numbers $|LM_L \rangle$ account for the relative
orientation of the collision pair in the lab frame.  Thus the
basis states of our calculation are given by
\begin{equation}
|s_d {\bar \lambda}{\bar \omega}\epsilon (j i_d) f_d m_{f_d} \rangle
|(s_ai_a)f_a m_{f_a} \rangle |LM_L \rangle.
\end{equation}
In this basis the total parity $p=\epsilon(-1)^{j-s_d+L}$ and the
lab-projection of total angular momentum, $m_d + m_a + M_L$, are
conserved quantities.  In zero field, the total angular momentum
${\cal F} = {\bf f}_d + {\bf f}_a + {\bf L}$ would also be
conserved; however, we anticipate considering the action of a field
on the collisions, which would mix different ${\cal F}$ values. We
therefore do not exploit this symmetry here.

Cast in this basis, the Schr\"{o}dinger equation for Rb-OH
collisions takes the standard form of a set of close-coupled
equations in the variable $R$.  To solve these, we propagate the
log-derivative matrix $Y \equiv \psi^{-1} d\psi/dR$ using a
variable-step version of Johnson's algorithm \cite{Johnson}. Because
of the multiple PES's and the inclusion of hyperfine degrees of
freedom, the total number of scattering channels required to solve
the complete problem including spin (in excess of 25,000) is
prohibitively large.

To overcome this obstacle, we exploit a kind of frame-transformation
procedure \cite{Fano}.  We identify a suitable radius $R=R_0$, and
define $R<R_0$ and $R>R_0$ as the ``short-range'' and ``long-range''
parts of the calculation, respectively. For Rb-OH, we choose
$R_0=17\ a_0$. In the short-range region, the PES's are deep, and
the hyperfine effects are small.  We therefore compute $Y$ in a
basis of decoupled nuclear spins and neglect couplings between
blocks of the Hamiltonian with different nuclear spin projections
$m_{ia}$ or $m_{id}$.  In this pilot study, we have neglected the
influence of the ion-pair channel.

At long range, $R>R_0$, the full hyperfine structure is restored,
and $Y$ is transformed into the basis (1) for further propagation
and matching to spherical Bessel functions to yield scattering
matrices. In this region, collision channels in which the molecule
is excited into higher-lying rotational or spin-orbit states are
already strongly closed. We therefore eliminate these channels. In
the inner region, we use partial waves up to $L=28$ and rotational
states up to $j=11/2$. In the outer region, we employ partial waves
up to $L=22$, but with full hyperfine structure of the $j=3/2$
rotational ground state.  Because of these approximations, no single
calculation requires more than 2000 channels \cite{fullpaper}.
Numerical checks suggest that the magnitudes of cross
sections computed in this way are accurate to within a factor of
$\sim 2$.

\begin{figure}
\centerline{\includegraphics[width=1.0\linewidth,height=0.8\linewidth,angle=0]{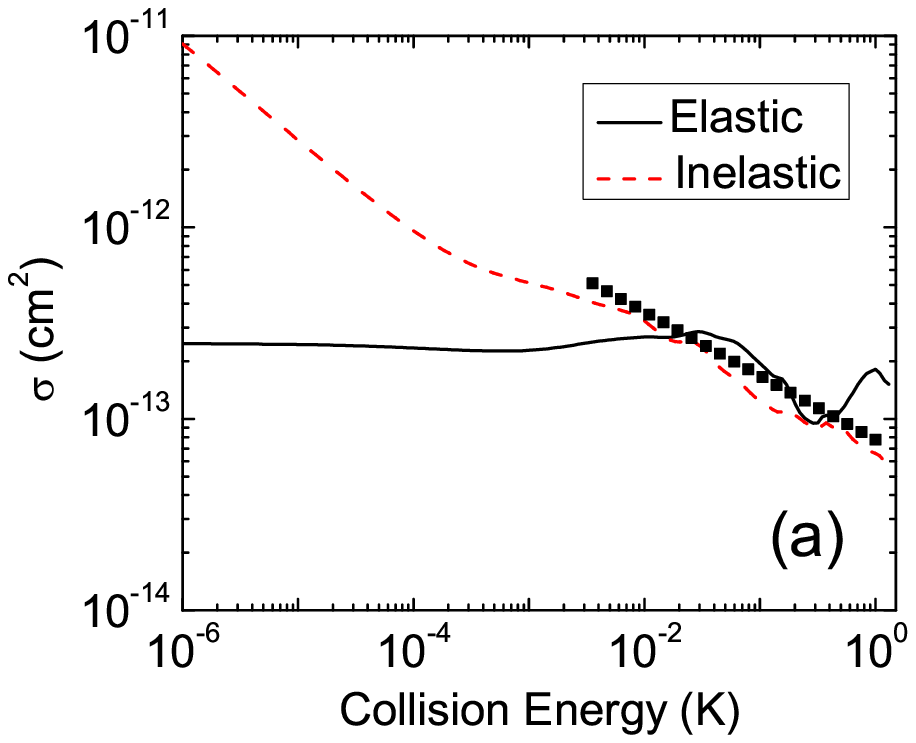}}
\centerline{\includegraphics[width=1.0\linewidth,height=0.8\linewidth,angle=0]
{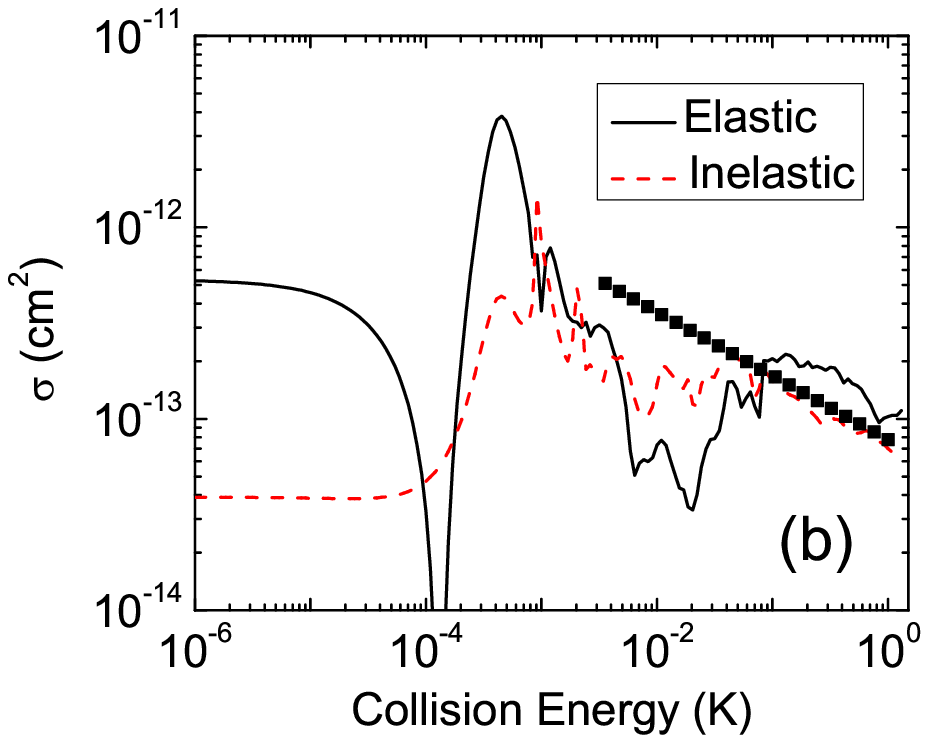}}
\caption{Rb-OH collision cross sections versus energy. Solid
(dashed) lines represent elastic (inelastic) cross sections. In (a)
is shown results for the incident channel $|^2\Pi_{3/2},
f_{d}m_{fd}=22, f \rangle_{\rm OH}$ $|f_{a}m_{fa} = 22 \rangle_{\rm
Rb}$, where both the atom and diatom are weak-magnetic-field
seeking, and the OH is also weak-electric-field seeking.  In (b) is
shown the incident channel $|^2\Pi_{3/2}, f_{d}m_{fd}=11, e
\rangle_{\rm OH}$ $|f_{a}m_{fa} = 1-1 \rangle_{\rm Rb}$.  Again,
both species are in weak-magnetic-field seeking states, but the OH
is now in a strong-electric-field seeking state. The points indicate
the Langevin cross section, Eqn. (2)} \label{cross_section}
\end{figure}

For concreteness, we calculate cross sections for a beam experiment
in which the incident wave vector is coincident with the laboratory
quantization axis.  Energy-dependent cross sections for two
different initial states of the collision partners are shown in Fig.
\ref{cross_section}. In the first case, Fig. \ref{cross_section}(a),
the Rb atom is initially in its ``stretched'' state, with
$|f_am_{f_a} \rangle = |22 \rangle$, and the molecule is in its
higher-lying, $f$ parity state, and also spin stretched, with
$|f_dm_{f_d} \rangle = |22 \rangle$.  Both these states are
weak-magnetic-field seeking, meaning that they can be trapped
magnetically. In addition, OH in this state can be trapped
electrostatically. Solid lines denote elastic cross sections
$\sigma_{\rm el}$, in which all internal quantum numbers are
retained after the collision, and dashed lines refer to inelastic
cross sections $\sigma_{\rm inel}$, which represents the sum of
partial cross sections to many possible outcomes distinct from the
initial channel (including $m_a$ and $m_j$-changing collisions that
do not release energy at zero field).

For collisions between S-state atoms, being in a stretched state
greatly reduces inelastic cross sections. This is simply because the
$m_{f_a}$ projection of one atom cannot be reduced at the expense of
raising $m_{f_a}$ of the other atom. Rather, to conserve angular
momentum, such a collision can occur only if the partial wave
projection $M_L$ can change, which requires an anisotropic
interaction. For two S-state atoms, the only anisotropic interaction
is the magnetic dipole interaction between valence electrons, which
is very weak. For Rb-OH collisions, by contrast, the PES itself is
highly anisotropic, providing a ready avenue for spin-changing
collisions. Indeed, inelastic scattering is as likely as, or more
likely than, elastic scattering over much of the energy range.  This
trend is especially pronounced at low energies (below $\sim 1$ mK),
where the Wigner threshold law sets in.  The elastic cross section
must tend to a constant at threshold, but exoergic inelastic
processes must yield cross sections that diverge as $1/\sqrt{E}$.
Note that the actual magnitudes of cross sections in this limit are
strongly subject to details of the PES, details which are typically
uncovered only by experiments.  Thus the cross sections in Fig.
\ref{cross_section}, while suggestive, cannot be taken as definitive
predictions.

At higher energies, many partial waves can contribute to the
scattering, and the threshold laws are less significant. Because the
electron spin  of the OH is strongly tied to the rotating molecule,
spin orientation in the lab frame is completely disrupted during a
collision.  This process can be approximated simply by a Langevin
model \cite{Levine}: for a given energy $E$, there is a maximum
partial wave $L(E)$ for which $E$ lies above the centrifugal barrier
of the long-range potential $-C_6/R^6 + \hbar^2 L(L+1)/2 \mu R^2$.
Here $C_6 = 325 E_h a_0^6$  is the isotropic Van der Waals
coefficient of the PES \cite{fullpaper}, and $\mu$ the reduced mass
of the collision pair. In the Langevin model, we assume that any
partial wave smaller than $L(E)$ contributes to inelastic scattering
with unit probability, i.e., any time the atom and molecule are near
one another, they are essentially guaranteed to disrupt the
electron's spin. This idea leads to a cross section
\begin{equation}
\sigma_{\rm Langevin}(E) = 3 \pi \left( \frac{C_6}{4E} \right)^{1/3}.
\label{Langevin}
\end{equation}
This cross section is also shown in Fig. \ref{cross_section}(a), by
the solid symbols.  It is clearly getting the trend and the order of
magnitude of the cross sections correct.

We therefore predict that Rb-OH elastic cross sections, at
temperatures of tens of mK typical of Stark decelerators, will be on
the order of $10^{-13}$ cm$^2$. At the corresponding collision
velocities of several hundred cm/s, this corresponds to elastic
collision rate constants on the order of $10^{-11}$ cm$^3$/s, which
is easily large enough to provide sufficient thermal contact to
allow sympathetic cooling in overlapped traps of OH and Rb. However,
inelastic rate constants are of the same order of magnitude, which
bodes ill for sympathetic cooling, at least in magnetostatic or
electrostatic traps.  Including the ion-pair channels would likely
only disrupt the internal states more severely, and would not
change this basic conclusion about sympathetic cooling.

We turn now to an alternative pair of initial states, whose
scattering cross sections are shown in Fig. \ref{cross_section} (b).
Here the atom is in its lowest magnetically trappable state, with
$|f_am_{f_a} \rangle = |1 -1 \rangle$, and the molecule is in its
lower-energy $e$ state, with total spin $|f_dm_{f_d} \rangle = |11
\rangle$ (also magnetically trappable). At zero field, both
$\sigma_{\rm el}$ and $\sigma_{\rm inel}$ are energy-independent at
low energies, since all exit channels are iso-energetic with the
incident channel. In nonzero field, the $|11 \rangle_{\rm OH} |1-1
\rangle_{\rm Rb}$ channel lies above other $m$-components in energy,
and $\sigma_{\rm inel}$ would again diverge.

By the same token, 
sympathetic cooling is possible at low energies for OH molecules in the $|1-1
\rangle$ state and Rb atoms in the $|11 \rangle$ state. This is the
lowest-energy channel of all: there are no
inelastic channels energetically available and $\sigma_{\rm inel}$
will vanish. These states are not magnetically trappable, but could
be confined in an optical or microwave \cite{DeMille} dipole trap or
an alternating current trap \cite{vanveld}. A small static magnetic
field will be needed to maintain the projection quantum numbers.

At higher collision energies, other forms of inelastic scattering
become energetically allowed.  Once the collision energy surpasses
the height of the $p$-wave centrifugal barrier (at about 1.6 mK),
the molecule can shed angular momentum into the
partial wave degree of freedom.  At this point $\sigma_{\rm inel}$
climbs to a large value, comparable to $\sigma_{\rm el}$.  Further,
at about 4mK, the $f=2$ hyperfine state of OH becomes energetically
allowed, providing another route to inelastic collisions.  By
examining the partial cross sections, we find that these two avenues
for inelastic collisions are roughly equally likely.  At still
higher collision energies, $\sigma_{\rm inel}$ is again roughly
approximated by the Langevin result (\ref{Langevin}), although
there is now more structure, owing to a large number of Feshbach
resonances to fine-and hyperfine excited states.

These results present a cautionary tale for sympathetic cooling
using alkali atoms.  One may regard inelastic scattering as a
``Murphy's Law'' process \cite{Matthews}. Namely, if the internal
state of the molecule can change to produce an unfavorable result,
it will do so. The key to making sympathetic cooling viable lies in
eliminating undesirable channels as far as possible. For example, a
light collision partner with a small $C_6$ coefficient will produce
a centrifugal barrier at higher energy, preventing partial waves
from accepting angular momentum.  This circumstance explains, at
least partly, the ability of a He buffer gas to cool molecules
without badly disrupting spin orientation
\cite{Weinstein,Bohn,Krems03}. In the present context of alkali
atoms, consider Li, which is about 12 times lighter than Rb, and
about half as polarizable.  Its $p$-wave centrifugal barrier upon
colliding with OH is on the order of  10 mK. In addition, a
molecule that is better described by Hund's coupling case (b), in
which the electron's spin is only weakly coupled to the molecular
rotation axis, may help weaken $\sigma_{\rm inel}$ below the
Langevin limit \cite{Krems03}.

In summary, we have performed the first {\it ab initio} scattering
calculations for an open-shell, ground state molecule
colliding with an alkali atom, incorporating the
hyperfine structure of both collision partners.  The results
suggest that elastic cross sections are sufficiently large
for sympathetic cooling to occur, yet equally large inelastic
cross sections probably hinder this application.  Nonetheless,
the results do not preclude the possibility that, by applying
electric or magnetic fields, inelastic cross sections could
be suppressed \cite{Ticknor,Krems}.
This will be the subject of a future study.

ML and JLB gratefully acknowledge the NSF and the W. M. Keck
Foundation, PS acknowledges the M\v{S}MT \v{C}R (grant No. LC06002).

\end{document}